\documentclass[12pt]{article}
\usepackage{amsmath, amssymb}
\usepackage{graphicx}
\usepackage{epsfig}
\usepackage{graphicx}
\usepackage{verbatim}

\usepackage{color}
\usepackage{hyperref}

\usepackage{amsfonts}

\def\hybrid{\topmargin -20pt    \oddsidemargin 0pt
        \headheight 0pt \headsep 0pt
        \textwidth 6.25in       
        \textheight 9.5in       
        \marginparwidth .875in
        \parskip 5pt plus 1pt   \jot = 1.5ex}

\def\be{\begin{equation}}
\def\ee{\end{equation}}
\def\ba{\begin{eqnarray}}
\def\ea{\end{eqnarray}}

\def\del{\partial}
\newcommand{\eqn}[1]{(\ref{#1})}

\definecolor{orange}{rgb}{1,0.5,0}
\definecolor{green}{rgb}{0,1,0}
\definecolor{purple}{rgb}{.5,0,1}

\def\a{\alpha}

\def\b{\beta}

\def\g{\gamma}
\def\G{\Gamma}

\def\D{\Delta}

\def\p{\pi}

\def\th{\theta}

\def\m{\mu}
\def\n{\nu}
\def\om{\omega}
\def\Om{\Omega}
\def\l{\lambda}

\def\s{\sigma}

\definecolor{orange}{rgb}{1,0.5,0}
\definecolor{green}{rgb}{0,1,0}
\definecolor{purple}{rgb}{.5,0,1}

\hybrid

\def\baselinestretch{1.2}

\catcode`\@=11

\def\marginnote#1{}
\begin{document}


\begin{titlepage}
\hfill
\vskip 2cm

\begin{center}

{ \Large
{\textbf
{Brane embeddings in sphere submanifolds}
}}

\vskip 0.9in
{\large {\bf Nikos Karaiskos}, {\bf Konstadinos Sfetsos}\ and { \bf Efstratios Tsatis}}
\vskip 0.2in
{Department of Engineering Sciences, University of Patras \\
26110 Patras, Greece\\
\footnotesize{\tt $\{$nkaraiskos, sfetsos, etsatis$\}$@upatras.gr}}\\

\end{center}
\vskip 1.0 cm

\centerline{\bf Synopsis}

\noindent Wrapping a D(8-p)-brane on $AdS_2$ times a submanifold of
$S^{8-p}$ introduces point-like defects in the context of AdS/CFT
correspondence for a Dp-brane background. We classify and work out the details in
all possible cases with a single embedding angular coordinate. Brane
embeddings of the temperature and beta-deformed near horizon
D3-brane backgrounds are also examined. We demonstrate the relevance
of our results to holographic lattices and dimers.

\end{titlepage}
\newpage

\def\baselinestretch{1.2}
\baselineskip 20 pt


\tableofcontents

\section{Prolegomena}

When branes wrap internal manifolds, they have the tendency to shrink. However, they can be
stabilized by turning on a worldvolume gauge field with a quantized flux.
A pioneering example of such flux stabilization of D-branes
was presented in \cite{Bachas:2000ik} for the wrapping of a probe D2-brane inside an $SU(2)$ group manifold
and a connection with results from an exact CFT approach was also made.
There have been numerous works in the literature considering brane embeddings in
various backgrounds and dimensions
\cite{Bachas:2000fr}-\cite{Skenderis:2002vf} (and references therein).
In particular, the authors of \cite{Camino:2001at} considered configurations where a D(8-p)-background
wraps an $\mathbb{S}^{7-p}$ inside an $\mathbb{S}^{8-p}$ in the background of a Dp-brane.
The stabilization occurs at quantized values of the equatorial angle of the
bigger sphere.
This result, besides being aesthetically beautiful, is also relevant in
a holographic approach to condensed matter lattices and dimer systems 
\cite{Kachru:2009xf}.

\noindent
Motivated by these works we realized that for generic values of $p$ there are 
more embedding possibilities even when one considers the
simplest case of one embedding coordinate.
Instead of $\mathbb{S}^{7-p}$, one could also
select other submanifolds of $\mathbb{S}^{8-p}$ whose
isometry groups are essentially given by subgroups of $SO(9-p)$, the latter being the isometry
group of $\mathbb{S}^{8-p}$. These submanifolds are presented in Table 1 for $p=0,1,\dots, 5$.
The coloring is introduced for later convenience.

\begin{table}[h]
 \caption{Submanifolds  of $\mathbb{S}^{8-p}$. }

\begin{center}
\begin{tabular}{|c|c|}
\hline
$p = 0$ & $\mathbf{S^7}$,
~~{\color{blue}$\mathbf{S^3 \times S^4}$},
~~{\color{purple}$\mathbf{S^2 \times S^5}$},
~~{\color{red}$\mathbf{S^1 \times S^6}$} \cr
\hline
$p = 1$ & $\mathbf{S^6}$,
~~{\color{blue}$\mathbf{S^3 \times S^3}$},
~~{\color{purple}$\mathbf{S^2 \times S^4}$}, ~~
{\color{red}$\mathbf{S^1 \times S^5}$},
~~{\color{red}$\mathbf{CP^3}$} \cr
\hline
$p = 2$ & $\mathbf{S^5}$,
~~{\color{blue}$\mathbf{S^2 \times S^3}$},
~~{\color{red}$\mathbf{S^1 \times S^4}$} \cr
\hline
$p = 3$ & $\mathbf{S^4}$,
~~{\color{blue}$\mathbf{S^2 \times S^2}$},
~~{\color{red}$\mathbf{S^1 \times S^3}$},
~~{\color{red}$\mathbf{CP^2}$} \cr
\hline
$p = 4$ & $\mathbf{S^3}$, ~~{\color{red}$\mathbf{S^1 \times S^2}$} \cr
\hline
$p = 5$ & $\mathbf{S^2}$, ~~{\color{red}$\mathbf{S^1 \times S^1}$} \cr
\hline
\end{tabular}
\end{center}

\end{table}
\noindent
This paper is organized as follows:
In section 2 we minimize the action of the brane probe and calculate the semi-classical
energy for each one of the aforementioned configurations. In general, the energy
depends on the ratio of the flux units $n$ of the worldvolume gauge field
to the number of the Dp-branes $N$, that we stack together to form the background.
For a given value of $p$ these energies depend on the specific submanifold that is wrapped.

\noindent
In section 3 we present brane embeddings in $\b$-deformed backgrounds \cite{Lunin:2005jy}.
In this case it turns out that the $\g$ dependence of the deformation drops out
completely in the probe computation. Pertaining
to the $\s$-deformation, which involves an S-duality, we formulated the problem mathematically,
but we were not able to find minimal configurations explicitly due to its complexity.

\noindent
In section 4, we turn on the temperature and examine its
effect on the stability of our constant embeddings. We conclude, by
considering a small fluctuation analysis, that these are
perturbatively stable.

\noindent
In section 5 we apply our results in the context of holographic lattices and dimers. 
We show that the free energy, and hence the physical behavior of the systems, is 
sensitive in a simple manner to the different wrappings we have constructed.
Finally, in section 6 we present concluding remarks and comment 
on future directions.

\section{Brane embeddings in Ramond-Ramond backgrounds}
The geometry created by a stack of $N$ coincident Dp-branes in the near-horizon
region is described by the ten-dimensional metric \cite{Horowitz:1991cd}
\be
ds^2 = \left(\frac{r}{R}\right)^{\frac{7-p}{2}} (-dt^2 + d\vec{x}^2_{||} )
+ \left(\frac{R}{r}\right)^{\frac{7-p}{2}} (dr^2 + r^2 d\Om_{8-p}^2)\ ,
\label{10dim_metric}
\ee
where $d\Om_{p}^2$ is generally the
line element of a unit $p$-sphere and the parameter $R$ is given by
\be
R^{7-p} = N ~ g_s ~ 2^{5-p}~ \pi^{\frac{5-p}{2}}~ (\a')^{\frac{7-p}{2}}~
\G(\tfrac{7-p}{2})\ .
\label{R_def}
\ee
The background is also supported by a dilaton, $\Phi(r)$ and a non-zero
Ramond--Ramond (RR) field strength $F_{(8-p)}$ given by
\ba
e^{-\Phi(r)} & = & \left(\frac{R}{r}\right)^{\frac{(7-p)(p-3)}{4}} \ ,
\cr
F_{(8-p)} & = & (7-p)R^{7-p}~\textrm{Vol}(\mathbb{S}^{(8-p)}) = \textrm{d}C_{(7-p)}\ ,
\label{dil_RR}
\ea
where Vol$(\mathbb{S}^{(8-p)})$ denotes the volume form of the unit $p$-sphere
and $C_{(7-p)}$ is the RR potential. We split the $(8-p)$ spherical coordinates as
$(\th, \phi_1, \dots , \phi_{7-p})$ and let $\th$ and $\vec{x}_{\parallel}$ be 
the embedding coordinates of the probe brane.

\noindent
We concentrate first to the cases corresponding to the entries of the Table 1 that involve solely spheres.
For these cases the metric of the compact space will have the form
\be
d\Om_{8-p}^2 = d\th^2 + \cos^2\th ~d\Om_q^2 + \sin^2\th ~d\Om_{7-p-q}^2\ ,
\qquad q=0,1,\dots , {\left[7-p\over 2\right]}\ .
\label{metric_sph}
\ee
This parametrization of the metric corresponds to splitting the ${\bf 9-p}$ representation
of the symmetry group $SO(9-p)$ under the subgroup $SO(q+1)\times SO(8-p-q)$ as
${\bf (9-p)\to (q+1,1)\oplus (1,8-p-q)}$. The variable $\displaystyle \th\in [0,{\pi\over 2}]$,
unless $q=0$ in
which case $\th\in [0,\pi]$. Consequently, the RR potential will be written as
\be
C_{(7-p)}  =  R^{7-p}~f(\th)~ \om_{(7-p)}\ ,
\label{RR_pot}
\ee
where
\be
\om_{(7-p)} = d{\rm Vol}(\mathbb{S}^q) \wedge d{\rm Vol}(\mathbb{S}^{7-p-q})
=\sqrt{h} d\phi_1\wedge d\phi_2 \dots d\phi_{7-p}\ .
\ee
The corresponding volume is
\be
\mathbb{V}_{\mathcal{M}}
= \int_{\mathcal{M}} d^{7-p}\phi\sqrt{h}
= \frac{4\pi^{\frac{9-p}{2}}}{\G\left(\frac{1+q}{2}\right)
\G\left(\frac{8-p-q}{2}\right)}\ .
\label{vmsn}
\ee
For $q=0$ we should divide this formula by two since the general expression for
${\rm Vol}(\mathbb{S}^q)$ gives $2$ for $q=0$.
The function $f(\th)$ is given for $q\neq 0 $ by
\be
f(\th)  =  \frac{7-p}{8-p-q} (\sin\th)^{8-p-q} ~{}_2F_1\left(\frac{1-q}{2},
4-\frac{p+q}{2},5-\frac{p+q}{2},\sin^2\th\right) \ ,
\label{RR_pot1}
\ee
and for $q=0$ by
\be
f(\th) = 2^{8-p}\frac{7-p}{8-p} \left(\sin\tfrac{\th}{2}\right)^{8-p} ~
{}_2F_1\left(\frac{p}{2}-3,4-\frac{p}{2},5-\frac{p}{2},\sin^2\tfrac{\th}{2}\right) \ .
\ee
This difference originates from the two different ranges that the angular variable 
$\th$ takes, as mentioned above.
Note also that this is not the most general form for
the RR potential, but it is the only one consistent for the particular embedding
that we will consider in this article.

\noindent
The D(8-p)-brane probe is described by the sum of a Dirac-Born-Infeld (DBI)
and a Wess-Zumino (WZ) term
\be
S = -T_{8-p}\int d^{9-p}\s e^{-\Phi} \sqrt{-\textrm{det}(\hat{g} + F)}
+ T_{8-p} \int C_{(7-p)} \wedge F\ ,
\label{probe_act}
\ee
where $\hat{g}$ is the induced metric on the brane, $F$ is an abelian gauge field
strength living on the world-volume of the brane and
\be
T_{8-p} = (2\pi)^{p-8}~ (\a')^{\frac{p-9}{2}}~ (g_s)^{-1}\ ,
\ee
is the tension of the probe brane.
The integration is performed over the world-volume coordinates of the brane
which are taken to be $\s^\a = (t, r,\phi_1, \dots, \phi_{7-p})$. In general,
the embedding coordinate may depend on any world-volume coordinate. Here, we
shall restrict ourselves to the case where $\th$ depends only on the radial coordinate,
which is also consistent with the form of the RR potential \eqn{RR_pot}. Since
the WZ term acts as a source term for the abelian gauge field strength $F$,
the latter one is constrained to be $F = F_{tr} ~dt\wedge dr$. We also set
the spacelike worldvolume coordinates $\vec{x}_{||}$
to constants which is consistent with their equations of motion.

\noindent
Given the above conditions the probe brane action assumes in general the form
\be
S = \int_\mathcal{M} d^{7-p}\phi \int dtdr \mathcal{L}(\th, F)\ ,
\ee
where the Lagrangian density is computed to be
\be
\mathcal{L}(\th,F) = -T_{8-p}R^{7-p}\sqrt{h}\left[\frac{f'(\th)}{7-p}
\sqrt{1-F_{tr}^2 + r^2\th'^2} - f(\th)F_{tr} \right].
\label{Lagrangian}
\ee
By varying the Lagrangian density with respect to the
worldvolume gauge potentials $A_t$ and $A_r$, one observes that 
\be
\frac{\del\mathcal{L}}{\del F_{tr}} = {\rm const}\ .
\ee
Then, it turns out that the gauge field assumes the form
\be
F_{tr} = \frac{f''(\th)}{\sqrt{(7-p)^2(f'(\th))^2 + (f''(\th))^2}}\ .
\ee
In order to attribute physical meaning to this constant, we consider the coupling
of our system to fundamental strings \cite{Camino:2001at}. This is achieved by replacing
$F$ with $F-B$ in $\mathcal{L}$, where $B$ is the Kalb--Ramond field. By expanding, at 
first order in $B$ we pick out a term of the form
\be
\int_{\mathcal{M}} d^{7-p}\phi  \int dt dr ~\frac{\del L}{\del F_{tr}}  ~B_{tr}\ .
\ee
We can interpret the coefficient in front of $B_{tr}$ as a charge ($n$ units of $T_f$)
that multiplies the Kalb-Ramond potential of the fundamental string. Therefore,
the fundamental strings ``feel'' a potential in this background, whose strength
is proportional to their number, $n$, and their tension $T_f = 1/(2\p \a')$.
Consequently, one writes
\be
\int_\mathcal{M} d^{7-p}\phi ~\frac{\del\mathcal{L}}{\del F_{tr}} = nT_f\ ,
\qquad n\in \mathbb{Z}\ .
\label{quant_cond}
\ee
In order to find semi-classical minima of the embeddings, solving the
equations of motion arising from the Lagrangian density would suffice.
However, since we are also interested in computing the energies of our
configurations, we will obtain the minima through the Hamiltonian
procedure. By performing a Legendre transformation, which
actually removes the WZ part, the Hamiltonian of
the system is given by
\be
H = \int_\mathcal{M} d^{7-p}\phi \int dtdr \left[\frac{\del\mathcal{L}}
{\del F_{tr}}F_{tr} - \mathcal{L} \right]\ .
\label{Hamiltonian}
\ee
Using the explicit form of the Lagrangian \eqn{Lagrangian} and the quantization
condition \eqn{quant_cond} the Hamiltonian becomes
\be
H = \l N T_f  \int dtdr \sqrt{1 + r^2\th'^2}
\sqrt{\left(\frac{f'(\th)}{7-p}\right)^2 +  \Big(\n \l^{-1} - f(\th) \Big)^2}\ ,
\label{hammil}
\ee
where we have defined
\be
\n = {n\over N} \ ,\qquad \l  = \frac{\G\left(\tfrac{7-p}{2}\right)}{\G\left(\tfrac{8-p-q}{2}\right)
\G\left(\frac{1+q}{2}\right)}\ .
\ee
Since the origin of the constant $\l$ is $\mathbb{V}_{\mathcal{M}}$, it turns out that, 
for reasons
explained below \eqn{vmsn}, for $q=0$ we should divide the above formula by two.
It is obvious from the expression for $H$ that it is
consistent to look for constant $\th$ configurations, since in this case the $r$-dependence
drops out. Setting $\th'=0$ and requiring $\del H/\del \th =0$ gives the condition
\be
f''(\th) = (7-p)^2(\n \l^{-1} - f(\th))\ .
\label{minima_eq}
\ee
As one can see in Table 2 in some cases, depending on the specific
values for $p$ and $q$, this equation admits an exact solution $\th(\l)$,
but in general it can only be solved numerically.
In the rest of the paper,
in order to avoid a plethora of symbols, we will denote by $\th$ the solution of 
\eqn{minima_eq}. The energy density is defined by
\be
H = \int dr ~\mathcal{E}\ .
\ee
For general values of $p$ and $q$ it is given by
\be
\mathcal{E}_{p,q} = \l NT_f \sqrt{(\cos\th )^{2q}
(\sin\th)^{2(7-p-q)} +  \left(\n \l^{-1} - f(\th)\right)^2{}}\ .
\ee
For the case where a
one-cycle is manifest, that is $q=1$, the above formula as well as
the one for the minima have a much simpler form given by
\be
\sin\th = \left(\tfrac{7 - p}{6 - p}~\n\right)^{\frac{1}{5-p}},
\qquad \mathcal{E}_{p,1} = NT_f\sqrt{\n^2 + (\sin\th)^{12-2 p} - 2\n (\sin\th)^{7-p}}.
\label{pq1}
\ee
Noting that $f(0)=0$ and $f({\pi\over 2})=\l^{-1}$, we find the limiting behaviors
\be
\mathcal{E}_{p,q}= n T_f + {\mathcal O}(\n^2) ,\qquad \mathcal{E}_{p,q}=(N-n) T_f + {\mathcal O}(1-\n)^2\ .
\ee
The results of our computations regarding the minima and the corresponding
energies are summarized in the Table 2 below. In all cases, the angle $\th$ ranges
from 0 to $\pi/2$. We have also included two more cases, apart from the products
of spheres, which arise for odd values of $p$, by writing the $\mathbb{S}^{8-p}$
as a $U(1)$ bundle over $\mathbb{CP}^{\frac{7-p}{2}}$. We use the conventions
of \cite{Pope:1980ub} and \cite{Nilsson:1984bj} for the $\mathbb{CP}^2$ and
$\mathbb{CP}^3$, respectively.\footnote{Our embedding coordinate $\th(r)$ in these
cases is identified with the coordinates $\chi$ and $\m$ in equations (5)
and (4.1) in the references \cite{Pope:1980ub} and \cite{Nilsson:1984bj},
respectively.}  The normalizations for the metrics are such that
$R_{\m\n} = \frac{16}{p+1} ~g_{\m\n}$ (for the values $p=1$ and $p=3$ that are of interest to us).

\begin{table}[h]
 \caption{Minima and energies}

\begin{center}
\begin{tabular}{|c|c|c|c|}
\hline
-- & Cycles & Algebraic equations for minima
& $\mathcal{E}_{p,q}$ in units of $NT_f$\cr
\hline \hline
& $\mathbb{S}^3 \times \mathbb{S}^4$
&  $-90 \sin\th + 25 \sin3\th + 3 \sin5\th = -112\n$
& $\mathcal{E}_{0,3}$ \cr
$p = 0$ & $\mathbb{S}^2 \times \mathbb{S}^5$
& $450 \cos\th + 25 \cos 3\th - 27 \cos 5\th = 448(1-\n)$
& $\mathcal{E}_{0,2}$ \cr
&  $\mathbb{S}^1 \times \mathbb{S}^6$
& $\sin\th =  \left(\frac{7}{6}\n\right)^{1/5}$
& $\n\sqrt{1-\frac{35}{36}\left(\frac{7}{6}\n\right)^{2/5}}$ \cr
\hline
& $\mathbb{S}^3 \times \mathbb{S}^3$ &
$\sin\th = \n^{1/2} $
&  $\n(1-\n)$ \cr
$p = 1$ & $\mathbb{S}^2 \times \mathbb{S}^4$
&  $18\pi \n + 8\sin2\th + 5 \sin4\th = 36\th$ &  $\mathcal{E}_{1,2}$ \cr
&  $\mathbb{S}^1 \times \mathbb{S}^5$
& $\sin\th =  \left(\frac{6}{5}\n\right)^{1/4}$
& $\n\sqrt{1-\frac{24}{25}\left(\frac{6}{5}\n\right)^{1/2}}$ \cr
& $ \mathbb{CP}^3$ & as above & as above \cr
\hline
$p = 2$ & $\mathbb{S}^2 \times \mathbb{S}^3$ &
$\cos\th = \frac{-2^{1/3} + (5 - 5\n + \sqrt{27-50\n + 25\n^2})^{2/3}}
{2^{2/3}(5 - 5\n + \sqrt{27 - 50\n + 25\n^2})^{1/3}}
$
& $\mathcal{E}_{2,2}$ \cr
&  $\mathbb{S}^1 \times \mathbb{S}^4$
& $\sin\th = \left(\frac{5}{4}\n \right)^{1/3}$
& $ \n \sqrt{1-\frac{15}{32}\left(\frac{25}{2}\n^2\right)^{1/3}}$\cr
\hline
 & $\mathbb{S}^2 \times \mathbb{S}^2$
 & $\th = \frac{\pi}{2}\n$ & $\frac{1}{\pi}\sin \pi\n $ \cr
$p = 3$ & $\mathbb{S}^1 \times \mathbb{S}^3$
& $\sin\th = \left(\frac{4}{3}\n\right)^{1/2}$ &
$ \n \sqrt{1- \frac{32}{27}\n}$\cr
 & $\mathbb{CP}^2$ &  as above &  as above \cr
\hline
$p = 4$ & $\mathbb{S}^1 \times \mathbb{S}^2$ & $\sin\th = \frac{3}{2}\n$
& $ \n \sqrt{1-\frac{27}{16}\n^2}$\cr
\hline
$p = 5$ & $\mathbb{S}^1 \times \mathbb{S}^1$ & $\th=0$~ or~ $\th=\frac{\pi}{2}$
& singular solutions \cr
\hline
\end{tabular}
\end{center}

\end{table}

\noindent
We should clarify two subcases of the above table. Firstly, the results for the
$\mathbb{CP}^2$  and the $\mathbb{S}^1 \times \mathbb{S}^3$ submanifolds coincide.
This happens because the wrapping in the first case involves the $U(1)$ fiber
with group structure $\mathbb{S}^1$  and a submanifold inside $\mathbb{CP}^2$,
which has a similar structure with $\mathbb{S}^3$. The same happens with the
results for the $\mathbb{CP}^3$ and the $\mathbb{S}^1\times \mathbb{S}^5$
submanifolds. Secondly, for
$p=5$ we have the solutions $\th=0$ and $\th=\frac{\p}{2}$ which
correspond to the collapse of the D-brane at the poles of the 3-sphere, thus
rendering them singular.
We also note that we omit the respective equations that give the minima and energies for
the submanifolds $\mathbb{S}^{7-p}\subset \mathbb{S}^{8-p}$, which can be 
found in \cite{Camino:2001at}.

\noindent   Having obtained the energies for the various values for $p$, we plot them together
with the energies found in \cite{Camino:2001at} in the following five Figures. 
The colors (black, blue, purple and red) correspond to the entries with 
the same colors in Table 1. The energies are plotted as functions of
the ratio $\n$, in units of $N T_f$. Curves with the same value for $p$, but a different one for $q$, might
intersect. We also use the obvious notation $(q\perp q',\n)$. 

\noindent We observe from the figures below, that for a given value of $p$, the 
maximally symmetric submanifolds
corresponding to $q=0$ have the lowest energy.
When the submanifold in consideration
includes an $\mathbb{S}^1$ or a $\mathbb{CP}$ space, the ratio $\n$ cannot
exceed the  value $\displaystyle {6-p\over 7-p}$ found from \eqn{pq1}.
At this maximum value, the corresponding value of the energy
density is $\displaystyle {1\over 7-p}$.

\begin{figure}[h]
\begin{minipage}[b]{0.5\linewidth}
\centering
\includegraphics[scale=.6]{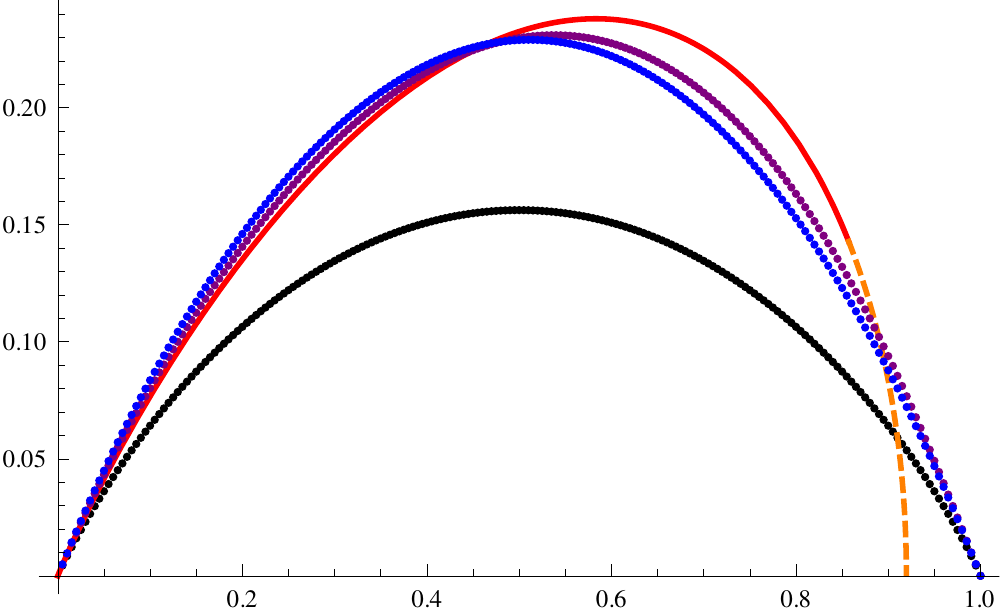}
\caption{Submanifolds for $p=0$. We have: ($1\!\perp\! 2$, $0.46$),
($1\! \perp\! 3$, $0.47$) and ($2\! \perp\! 3$, $0.48$).}

\label{fig:figure1}
\end{minipage}
\hspace{0.5cm}
\begin{minipage}[b]{0.5\linewidth}
\centering
\includegraphics[scale=.6]{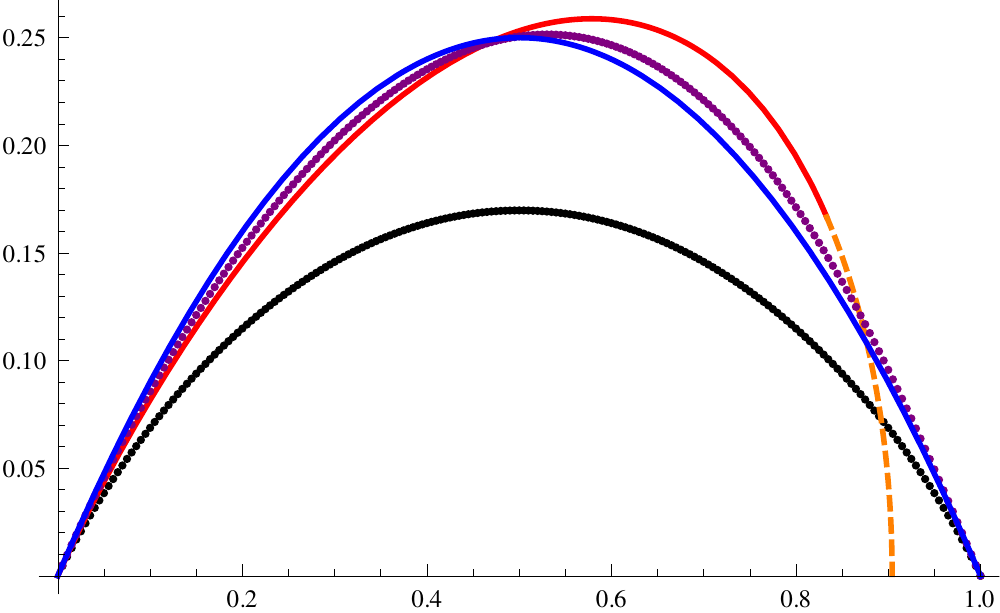}
\caption{Submanifolds for $p=1$. We have: ($1\!\perp\! 2$, $0.49$),
($1\! \perp\! 3$, $0.47$) and ($2\! \perp\! 3$, $0.51$).}
\label{fig:figure2}
\end{minipage}
\end{figure}

\begin{figure}[h]
\begin{minipage}[b]{0.5\linewidth}
\centering
\includegraphics[scale=.6]{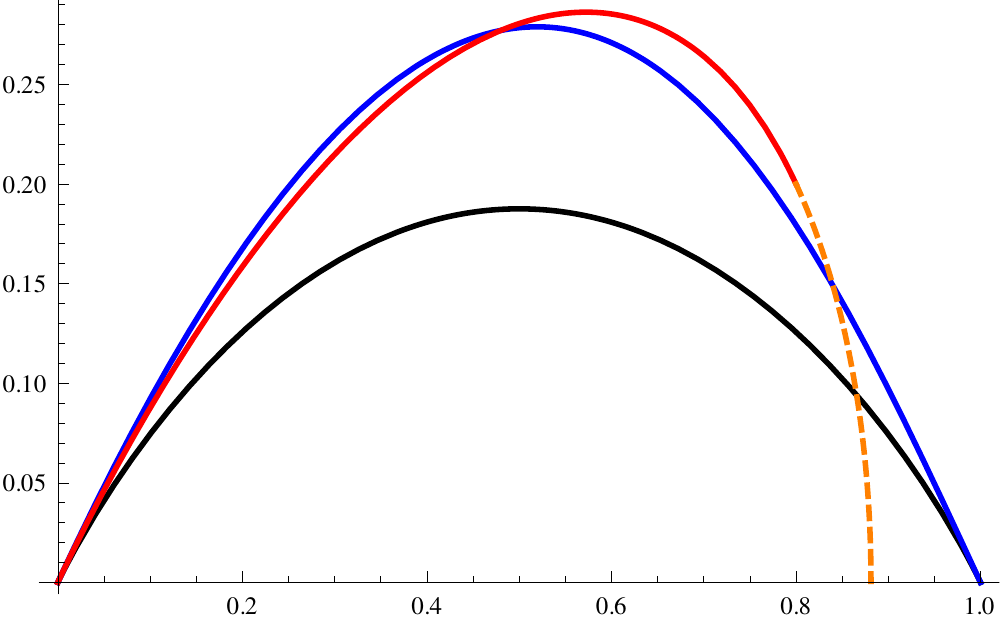}
\caption{Submanifolds for $p=2$. We have: ($1\! \perp\! 2$, $0.48$).}
\label{fig:figure3}
\end{minipage}
\hspace{0.5cm}
\begin{minipage}[b]{0.5\linewidth}
\centering
\includegraphics[scale=.6]{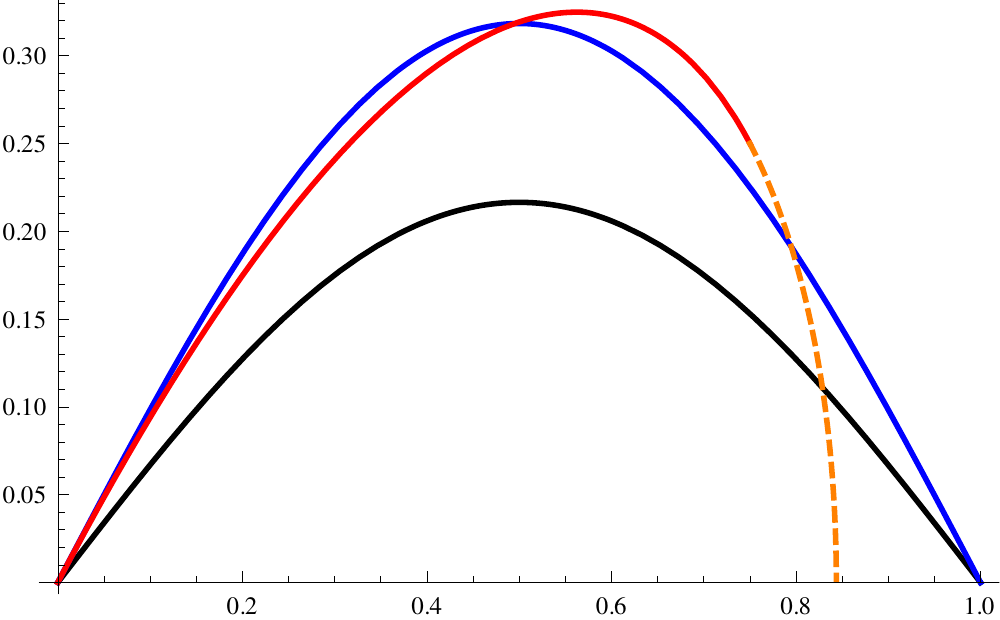}
\caption{Submanifolds for $p=3$. We have: ($1\! \perp\! 2$, $0.50$).}
\label{fig:figure4}
\end{minipage}
\end{figure}

\begin{figure}[!h]
\centering
\includegraphics[scale=.6]{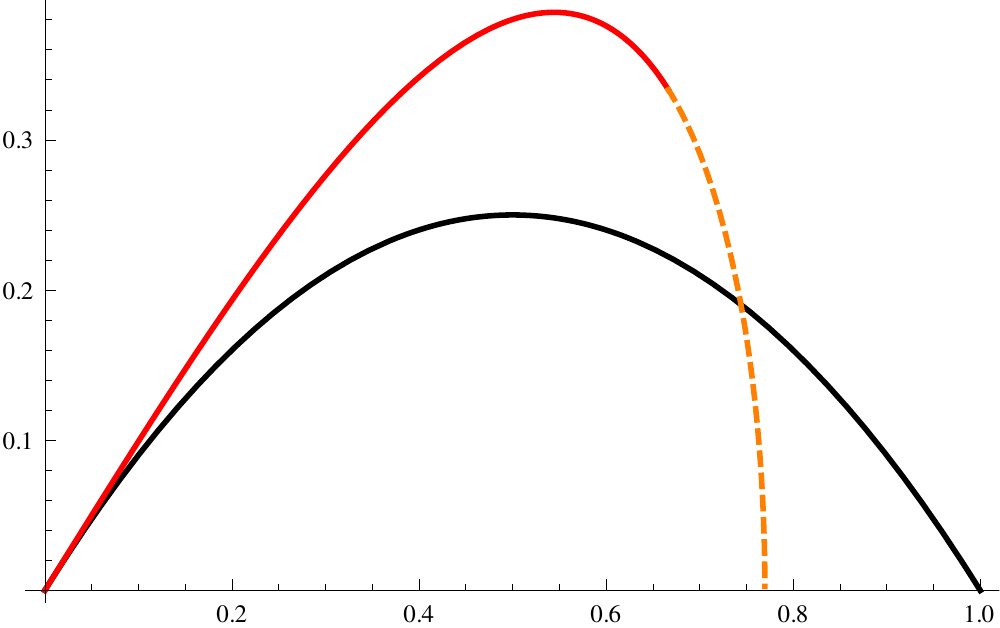}
\caption{Submanifolds for $p=4$.}
\label{fig:figure5}
\end{figure}

The stability of the configurations is ensured by performing a small fluctuation
analysis of the fields around the minima. We postpone the analysis until section 4,
where a finite temperature is also introduced.

\section{Brane embeddings in deformed backgrounds}
In this section we consider  brane embeddings inside $\b$-deformed background solutions 
of type-IIB  supergravity \cite{Lunin:2005jy}. We begin with
the $\g$-deformation of the $AdS_5 \times \mathbb{S}^5$ background for which the
$AdS_5$ part of the metric remains the same, while the metric of the
$\g$-deformed 5-sphere is written as
\be
d\Om_{5,\g}^2 = \sum_{i=1}^3 (d\m_i^2 + \mathcal{G} \m_i^2 d\phi_i^2)
+ \mathcal{G} R^4 \g^2 \m_1^2\m_2^2\m_3^2 (\sum_i d\phi_i)^2\ ,
\ee
where
\be
{\cal{G}}^{-1} = 1 + R^4\gamma^2 (\mu_1^2 \m_2^2 + \m_2^2 \m_3^2 + \m_3^2 \m_1^2)
\ee
and $(\m_1, \m_2, \m_3) \equiv (\cos\th, \sin\th\cos\psi, \sin\th\sin\psi)$.
The NS sector of the background includes a dilaton and a Kalb--Ramond
two-form, given by
\ba
e^{2\Phi} & = & \mathcal{G} e^{2\Phi_0}\cr
B_{NS} & = & \g R^4 \mathcal{G} (\m_1^2 \m_2^2 d\phi_1 \wedge d\phi_2
~~\textrm{+ cyclic})
\ea
and the RR potential and field strengths
\ba
C_2 & = & -4\g R^4 w_1\wedge (d\phi_1 + d\phi_2 + d\phi_3)\ ,
\qquad
\textrm{with} ~~ dw_1 = \tfrac{1}{2}\cos\th\sin^3\th\sin2\psi d\th\wedge d\psi\ ,
\cr
C_4 & = & 4R^4 (w_4 + \mathcal{G} w_1 \wedge d\phi_1 \wedge d\phi_2 \wedge d\phi_3)\ ,
\qquad \textrm{Vol}(AdS_5) = dw_4\ ,
\cr
F_5 & = & 4R^4 (\textrm{Vol}(AdS_5)+ \mathcal{G}~\textrm{Vol}(\mathbb{S}^5))\ ,
\qquad
\textrm{Vol}(\mathbb{S}^5) = dw_1 \wedge d\phi_1\wedge d\phi_2\wedge d\phi_3\ .
\ea
We consider D5-brane embeddings in this deformed background. The brane will
wrap the four angles of the deformed sphere so that the world-volume
coordinates will be $(t,r, \psi, \phi_i)$ and the embedding coordinates are taken
as $\vec{x}_{||} = {\rm const.}$ and $\th=\th(r)$. As before, we also turn on an Abelian world-volume
gauge field strength $F_{tr}$. The action of the brane probe
is given by a sum of a DBI and a WZ term. Some extra care is needed since
there are new terms arising from the induced Kalb--Ramond field and the RR potentials.
The action assumes the generic form
\be
S= - T_5 \int_{D5} e^{-\hat{\Phi}}\sqrt{P[g] + \cal{F}}
+ T_{5}\int_{D5}\sum_{p} C_{p}\wedge e^{\cal{F}}\ ,
\ee
where $\mathcal{F}$ is given by $\mathcal{F} = F -P[B]$, with $F$ and $P[B]$ being the 
world-volume field strength and the pullback of the Kalb-Ramond potential respectively. 
After performing the computation, the action 
for the D5-brane reads
\be
S = -\frac{T_5 R^4 }{2} \int d\psi d^3\phi\sin2\psi \int dtdr
(\cos\th\sin^3\th \sqrt{1 - F_{tr}^2 +r^2\th'^2} - F_{tr} \sin^4\th) \ .
\ee
The entire $\g$-dependence has dropped out completely due to non-trivial
cancellations in the DBI and WZ terms, separately. In fact, this action is exactly the same
as that computed for
the $p=3$ and $q=1$ case in which the D5-brane wraps the $\mathbb{S}^1 \times \mathbb{S}^3$
submanifold of $S^5$. Indeed, one may check that the above Lagrangian falls into the generic
family \eqn{Lagrangian} with $f(\th) = \sin^4\th$, which is the correct function
appearing in the RR potential for the aforementioned case.

\subsection{Embeddings in the $\s$-deformed background}

One may also consider a more general deformation of
the background, by performing an S-duality in the theory \cite{Lunin:2005jy}. Apart from $\g$, the
resulting background depends also on $\s$ which is an additional scaleless parameter.
Searching for D5-brane embeddings, we choose the embedding coordinates $\vec{x}_{||}={\rm const.}$ and
$\th=\th(\psi)$. As opposed to the previous cases, here $\th$ should depend on
$\psi$, since the latter enters in the computations in a non-trivial way.
Actually, as we shall explain later, this is related to the chosen embedding. The
Hamiltonian of the system turns out to be
\be
H = T_5 R^4 \sqrt{\mathcal{H}} \sqrt{P^2 Q + (P \sin^2\th - f(\psi))^2}\ ,
\ee
with
\ba
&& P = \frac{1}{2\mathcal{H}} \sin^2\th \sin2\psi \ , \qquad
Q  =  \mathcal{H}\sin^2\th - \sin^4\th + \mathcal{H} \th'^2\cos^2\th\ ,
\cr
&& ~f(\psi) = \frac{\n}{2}\sin2\psi\ , \qquad ~~~~
\mathcal{H} = 1 + \s^2R^4 (\mu_1^2 \m_2^2 + \m_2^2 \m_3^2 + \m_3^2 \m_1^2)\ ,
\ea
and the $\m_i$'s were defined in the previous section. As before,
the parameter $\g$ does not appear at all,
but $\s$ does. It should be obvious that an attempt to find constant
minima, namely $\psi$-independent solutions, is inconsistent. Varying the Hamiltonian
with respect to $\th$ gives a complicated nonlinear differential equation
that one has to solve in order to find configurations that minimize the energy.
We were unable to find solutions of this differential equation.

\noindent
This increased level of complexity occurs due to the particular embedding that
we considered. Had we chosen a similar embedding $\th = \th(\psi)$ for the
undeformed background, that is the $p=3$ case with manifest $\mathbb{S}^1 \times
\mathbb{S}^3$, 

\begin{figure}[ht]
\begin{minipage}[b]{0.5\linewidth}
\centering
\includegraphics[scale=.65]{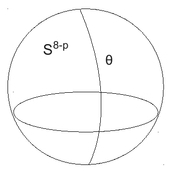}
\caption{Constant embedding}
\label{fig:figure6}
\end{minipage}
\hspace{0.5cm}
\begin{minipage}[b]{0.5\linewidth}
\centering
\includegraphics[scale=.65]{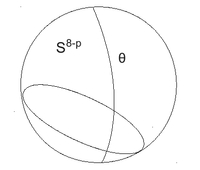}
\caption{Embeddings $\th=\th(\psi)$.}
\label{fig:figure7}
\end{minipage}
\end{figure}

\noindent
it would have also resulted to a similarly complicated differential
equation. The two
different embeddings in that case are depicted in Figures 6 and 7 above.

For non-constant embeddings $\th=\th(\psi)$ in the underformed case, it is obviously 
possible to rotate the north pole in a way to obtain the first configuration. In practice, 
this is done by performing an $SO(6)$ transformation. In the $\s$-deformed case
it is not obvious what the corresponding transformation would be, given that the isometry
group has been reduced.
The deformed sphere has
the same Euler characteristic with the undeformed one, so that their topology is
the same. It makes sense, then, to assume that such a transformation exists,
although we were not able to find it.

\section{Turning on temperature  }
It is natural to extend the discussion to asymptotically AdS spacetimes, which are
relevant to a holographic approach to dimers in condensed matter systems, as
pursued in \cite{Kachru:2009xf}. We will briefly discuss the near horizon geometry of
black D3-branes in such a way that the submanifold 
$\mathbb{S}^1 \times \mathbb{S}^3$ of $\mathbb{S}^5$ appears explicitly.
The metric of the background is
\be
ds^2=
-f(r)dt^2 + \frac{dr^2}{f(r)} + \frac{r^2}{R^2}d\vec{x}^2_{||} +
R^2 ( d\th^2  + \cos^2\th ~d\Om_1^2 + \sin^2\th ~d\Om_3^2)\ ,
\quad f(r) = \frac{r^4-\m^4}{R^2r^2}, 
\label{asympt_ads}
\ee
and the RR-potential changes also accordingly. The Hawking temperature is simply 
proportional to the parameter $\m$.
We consider a D5-brane probe
with the same embedding coordinates as before, i.e. $\vec{x}_{||}={\rm const.}$ and $\th=\th(r)$.
It is a straightforward task
to show that the minima of the particular configuration remain the same as with the
zero temperature case, this being true for every $p$. This wouldn't be the case
for more general $r$-dependent solutions.

\noindent
In order to ensure the stability of the configurations, one can consider small 
fluctuations around the minima. Let
\be
\th = \bar{\th} + \xi, \qquad F_{tr} = \bar{F}_{tr} + \chi\ ,
\ee
where the bars denote the minima and $\chi=\del_t \a_r - \del_r \a_t$. It should be stressed
out that this is consistent as long as one considers only the zero mode in the spherical
harmonic expansion on the $\mathbb{S}^5$. In order
to find the complete spectrum, one should also turn on fluctuations of the field
strength in every possible direction (see also \cite{Kruczenski:2003be} for a
prime example). However, here we are only interested in demonstrating perturbative 
stability at non-zero temperature, and for that, restricting to the zero-mode suffices.
The effective Lagrangian for quadratic fluctuations is found to be
\ba
\mathcal{L} & = & \tfrac{1}{2}\sqrt{h} R^4 T_5\cos\bar{\th}\sin^3\bar{\th}
\frac{1}{\sqrt{1 - \bar{F}^2_{tr}}}\left[
R^2\left(f(r)^{-1}(\del_t\xi)^2 - f(r)(\del_r\xi)^2\right)\right. \cr
  &  & \qquad\qquad +\left.\frac{\chi^2}{1 - \bar{F}^2_{tr}} + A\xi^2 + B\chi \xi\right]\ .
\label{s1s3lagr0}
\ea
The minima and the gauge field are given by
\be
\sin\bar{\th} = \sqrt{\frac{4}{3}\n }\ ,
\qquad \bar{F}_{tr} = \frac{9-16\n}{\sqrt{81-96\n}}\ ,
\label{minvals1s3}
\ee
and we have defined the constants
\be
A = 4 + \frac{36}{27 - 32 \n}\ ,
\qquad B = \sqrt{81-96 \n\over 3\n - 4 \n^2}\ .
\ee
We obtain the equations of motion by varying $\chi$ and $\xi$. After combining them
and concentrating on a Fourier mode of the form $\xi = e^{i\om t} \Psi(r)$,
we get
\be
{d\over dr}\left( f(r){d\Psi \over dr}\right) +\left({\om^2 \over f(r)} - {C\over 2 R^2}\right)\Psi(r)= 0\ ,
\qquad C \equiv 24+\frac{72}{32 \n - 27}\ ,
\ee
defined for $r\geqslant \m$. We transform this into a Schr\"odinger equation for 
$\Psi$, by appropriately changing
to a new variable $z=\int_r^\infty dr' f^{-1}(r')$, with $z\in [0,\infty)$ as
$r\in (\infty, \m]$.
The associated potential, that can be written explicitly only in terms or $r$, is
\be
V = \frac{C}{2}\frac{f(r)}{R^2}\ .
\ee
Substituting the value for $\th$ in $C$ from \eqn{minvals1s3}
one sees that $C$ is non-negative. Hence the zero mode of the configuration
is always positive. In fact, $C$
vanishes for the critical value $\n={3/4}$. In conclusion, the
configuration that we considered is stable. Similar arguments also hold for
the other submanifolds and for the cases $p =0,1,2,4$ as well.

\section{Application on holographic dimers}

It is interesting to investigate how the results of the previous sections affect the
holographic description of dimers. The main idea was pioneered
in \cite{Kachru:2009xf}. There, the authors considered lattices of D5-branes
embedded in a D3 black brane background in order to model a finite temperature system.
The chosen embedding is such that each probe brane wraps an $\mathbb{S}^4 \subset 
\mathbb{S}^5$. 
By generalizing the arguments presented there, we
will show in the present section that the less symmetric embeddings we found 
in section 2 are more favorable in the aforementioned context. 

\noindent
The metric of the background under consideration is \eqn{asympt_ads}, alongside
with the parametrization \eqn{metric_sph} for the compact manifold restricted
to the case with $p=3$. Hence the embedded branes wrap an $AdS_2$ space times a four dimensional submanifold. 
The free energy of the single D5-brane (or an anti-D5-brane) is computed by integrating the on-shell 
action \cite{Yamaguchi:2006tq}. In order to do this, one performs a Wick rotation 
to the Euclidean metric where time is identified as a periodic 
variable, which is the temperature. The free energy of a single D5-brane that goes straight 
down to the D3-horizon is given then by 
\be
F_{\textrm{D5}} = -\l \m N T_f \mathcal{E}_{3,q}\ .
\ee
We observe that the result is proportional to the energies that we computed in section 2. 
The computation is very similar to that performed for the $q=1$ case in
 \cite{Yamaguchi:2006tq} so that we omit the details. 

\noindent
As in \cite{Kachru:2009xf}, we will consider a lattice of D5- and an 
anti-D5-brane pairs. Each pair is essentially constituted by a D5-brane which
dives into the bulk and returns with an opposite orientation, thus regarded
as an anti-D5-brane. There exist two configurations then, for each pair,
depending on the value of the temperature. In the disconnected configuration
the D5- and the anti-D5-brane are separated and do not interact.\footnote{It 
turns out that all the essential details of analyzing this problem are 
similar to those in the holographic computation of the Wilson loop related to the 
binding energy of a quark-antiquark pair \cite{Rey:1998bq, Brandhuber:1998bs}.}
The total free energy of the pair is just the sum of the individual free energies,
that is simply equal to
\be
F_{\textrm{disconnected}} = 2F_{\textrm{D5}}\ .
\ee
As will become transparent below, this configuration dominates at high temperature.

\noindent
In the second configuration in which the D5- and the anti-D5-brane are connected 
with each other, the two membranes are separated by $\D x$ and are located 
at $r=\infty$ with $\vec{x} = (\pm \frac{\D x}{2},0,0)$. One considers then
embeddings with $\th(r)=\th_\n$ and $x=x(r)$. The turning point of the 
D5-brane is computed by $dr/dx' = 0$ and has the same form for a generic 
wrapping. We scale the turning point by the temperature and we define the dimensionless 
parameter $\displaystyle z_0 \equiv \frac{r_{\textrm{turn}}}{\m}$.
The turning point is associated with the spacing between the branes and the 
temperature by the following relation
\be
\frac{\m}{R^2}\D x = \left[ 2 (z_0^4 -1)^{1/2} \int_{z_0}^\infty
dz \sqrt{\frac{1}{(z^4-1)(z^4-z_0^4)}} \right]\ .
\ee
Noting the similarity with the holographic computation of the binding energy of 
a quark-antiquark pair we mentioned above, we perform
the integration obtaining \cite{Brandhuber:1999jr}
\be
\frac{\m}{R^2}\D x = \frac{1}{2} B(3/4,1/2) 
\frac{\sqrt{z_0^4-1}}{z_0^3} 
~{}_2F_1\left(\frac{1}{2}, \frac{3}{4}, \frac{5}{4}; 
\frac{1}{z_0^4} \right)\ .
\ee
As seen in Figure 8, for fixed lattice spacing $\D x$,
a solution of this type exists only for low enough temperatures. 
For higher temperatures, the disconnected configuration
is the only available solution. The critical temperature beyond which the latter 
dominates is not however given by the maximum value of the temperature, 
since the disconnected configuration already 
acquires a lower free energy at a lower temperature. 
To see that we compute the free energy of the connected configuration which is found to be
\ba
F_{\textrm{connected}} & = & 2 \l \m N T_f \mathcal{E}_{3,q} 
\left[- z_0  + \int_{z_0}^\infty dz 
\left(\sqrt{\frac{z^4 -1}{z^4 - z_0^4}}  - 1 \right) 
\right] \cr
&  \equiv  & 2 \l \m N T_f \mathcal{E}_{3,q} \tilde{F}_{\textrm{connected}}\ ,
\label{Fcon}
\ea
where we have defined the function
\be
\tilde{F}_{\textrm{connected}} = \frac{z_0}{4}B\left(-1/4,1/2 \right)
 {}_2F_1\left( -\frac{1}{2}, -\frac{1}{4}, \frac{1}{4} 
; \frac{1}{z_0^4} \right) \ ,
\label{fcontilde}
\ee
and we have computed the integral. This measures the deviation of the free energy 
from that of the disconnected configuration. To proceed with the analysis we note 
from Figure 8 that for the same temperature there exist two values of 
$z_0$. This mutlivalueness is also manifest in the plot of the free energy for the connected configuration in Figure 10.
Based on the experience with a general analysis for the quark-anti-quark binding 
energy performed in \cite{Avramis:2006nv}, we expect that 
a similar analysis here will indicate that the upper branch is unstable under small 
perturbations, so that we disregard it completely. Note also that in Figures 8, 9
and 10 the black, red and blue colored branches correspond to the unstable, 
meta-stable and stable branches, respectively.

\noindent Next, we compare the free energy of the connected configuration as a 
function of $z_0$ with that of the disconnected configuration. We see from Figure 9 
that for large values of $z_0$, equivalently for small temperatures, the connected 
configuration is more favorable. There exist a critical value of $z_0$, numerically 
equal to $z_{0,c} \simeq 1.52$,
\begin{figure}[ht]
\begin{minipage}[b]{0.5\linewidth}
\centering
\includegraphics[scale=.6]{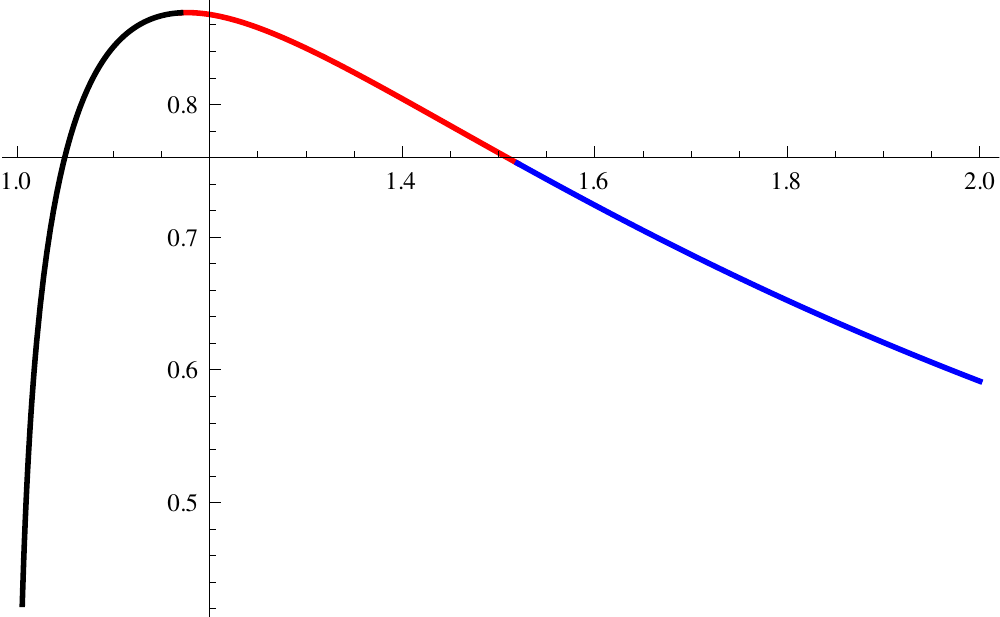}
\caption{$\frac{\m \D x}{R^2}$ as a function of $z_0$.}
\label{fig:figure8}
\end{minipage}
\hspace{0.5cm}
\begin{minipage}[b]{0.5\linewidth}
\centering
\includegraphics[scale=.6]{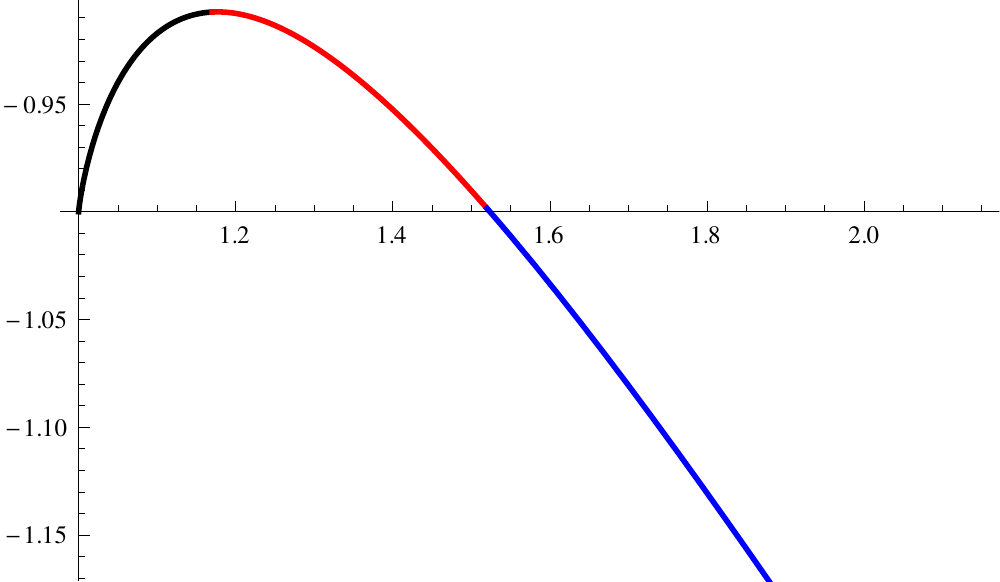}
\caption{$\tilde{F}_{\textrm{connected}}$ as a function of $z_0$.}
\label{fig:figure9}
\end{minipage}
\end{figure}
\noindent 
below which the disconnected configuration is favorable. Therefore, 
there exists a critical temperature at which the system undergoes a phase 
transition. This phase transition is of first
order, since the first derivative possesses a discontinuity at the critical
value $z_{0,c}$.
\begin{figure}[ht]
\centering
\includegraphics[scale=.6]{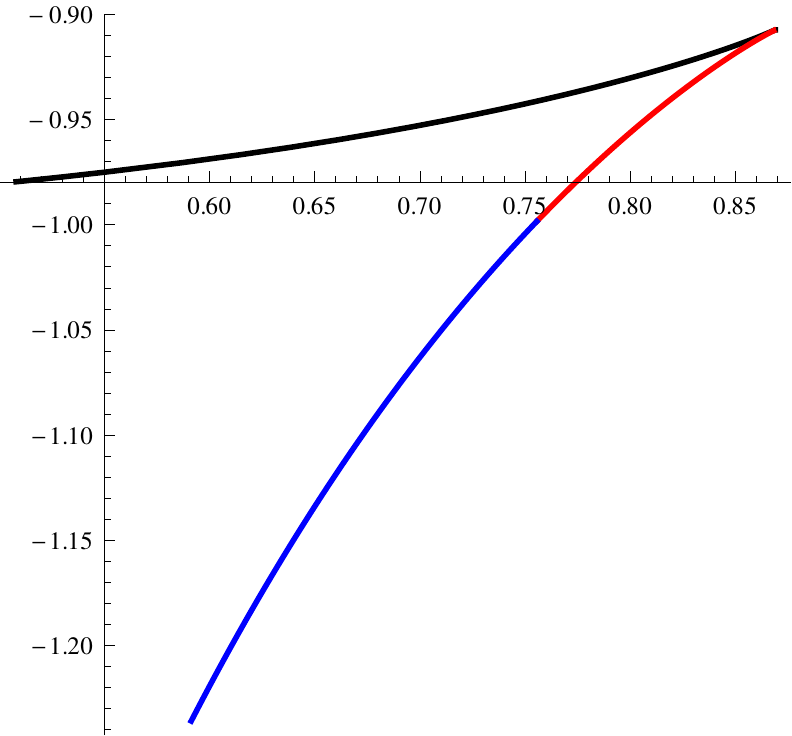} 
\caption{$\tilde F_{\textrm{connected}}$ as a function of $\frac{\m \D x}{R^2}$.}
\label{fig:figure10}
\end{figure}

\noindent
Eventually, in order to make contact with the results of section 2, we first observe 
that \eqn{fcontilde} is the same for all wrappings. Thus the only difference in
\eqn{Fcon} for different wrappings origins simply from the constant factor in 
front, which is essentially the energy of the wrapping. Since 
$\tilde{F}_{\textrm{connected}}$ is always negative then, the wrapping with 
the heighest energy density, which is less symmetric, 
has the minimal free energy, becoming more favorable in this context.
By considering lattices of pairs then one constructs dimers in a holographic 
way, along the lines presented in \cite{Kachru:2009xf}.

\section{Concluding remarks}

We classified and energetically compared all possible cases, with a single embedding angular coordinate,
in which
a D(8-p) brane can wrap $AdS_2$ times a submanifold of $\mathbb{S}^{8-p}$ in a Dp-brane background,
thus producing a pointlike defect. We worked out the details in all different
cases that arise, performing also comparisons between them.
We examined similar constructions in the presence of temperature and
in $\beta$-deformed backgrounds. We demonstrated stability by a small fluctuation 
analysis around the minima.

\noindent
It would be interesting to investigate and search for running solutions of the embedding 
coordinate, i.e. $\th=\th(r)$. This involves the classical equation of motion
for the Hamiltonian \eqn{hammil}. This is a highly non-linear equation but it should be 
possible to analyze it numerically. Of particular interest would be solutions connecting 
minima corresponding to different values of $n$, especially when they correspond to the 
same energy.

\noindent
Moreover, it would be very useful to extend our results beyond the probe approximation, 
by considering the backreaction of the probe branes on the background. This is 
significant when their number is comparable to the color number. 
In addition, these backreaction effects would also influence
the dimerization analysis presented here, in case where many branes are located at 
each lattice site.

\subsection*{Acknowledgements}

\noindent
We would like to thank K. Siampos and D. Zoakos for useful discussions on the subject.
N.K. acknowledges financial support provided by the Research Committee of the
University of Patras via a K. Karatheodori fellowship under contract number C. 915.

\end{document}